\documentclass[12pt]{iopart}
\usepackage{iopams}  
\usepackage{graphicx}

\begin{document}

\title{An ab initio based approach to optical properties of semiconductor heterostructures}

\author{L.~C.~Bannow$^1$, P.~Rosenow$^2$, P.~Springer$^1$\footnote{Current address: Institute of Technical Physics, German Aerospace Center, Paffenwaldring 38-40, 70569 Stuttgart, Germany}, E.~W.~Fischer$^2$, J.~Hader$^3$, J.~V.~Moloney$^3$, R.~Tonner$^2$, and S.~W.~Koch$^1$}

\address{$^1$ Department of Physics and Material Sciences Center, Philipps-Universit\"at Marburg, Renthof 5, 35032 Marburg, Germany}
\address{$^2$ Department of Chemistry and Material Sciences Center, Philipps-Universit\"at Marburg, Hans-Meerwein-Str. 4, 35032 Marburg, Germany}
\address{$^3$ Nonlinear Control Strategies Inc, 7040 N. Montecatina Dr., Tucson, AZ 85704, USA and College of Optical Sciences, University of Arizona, Tucson AZ 85721, USA}
\ead{\mailto{lars.bannow@physik.uni-marburg.de}}
\vspace{10pt}
\begin{indented}
\item[]March 2017
\end{indented}

\begin{abstract}
A procedure is presented that combines density functional theory computations of bulk semiconductor alloys with the semiconductor Bloch equations, in order to achieve an ab initio based prediction of the optical properties of semiconductor alloy heterostructures. The parameters of an eight-band $\mathbf{k\cdot p}$-Hamiltonian are fitted to the effective band structure of an appropriate alloy. The envelope function approach is applied to model the quantum well using the $\mathbf{k\cdot p}$-wave functions and eigenvalues as starting point for calculating the optical properties of the heterostructure. It is shown that Luttinger parameters derived from band structures computed with the TB09 density functional reproduce extrapolated values. The procedure is illustrated by computing the absorption spectra for a (AlGa)As/Ga(AsP)/(AlGa)As quantum well system with varying phosphide content in the active layer.
\end{abstract}

%
\vspace{2pc}

\submitto{\MSMSE}
%
%
%

\section{Introduction}
The majority of modern electronic devices are based on semiconductors. Therefore, a lot of research aims to improve their optical and electronic properties. Band gap engineering by means of semiconductor alloying has been proven a powerful tool for this purpose. Today, up to five different semiconductors are mixed resulting in quinary alloys, e.g. (GaIn)(NAsSb)~\cite{Li2003,Yuen2004,Harris2005,Jackrel2007,Harris2007} or (AlInGa)(AsSb)~\cite{Bugge2006,Hosoda2008,Suchalkin2008}. In addition, modern growth techniques allow the production of very pure compounds which is essential for optical properties. With this ever increasing number of available materials, it becomes increasingly important to predict the optical properties of new compounds to determine their usability for application in opto-electronic devices such as semiconductor lasers or solar cells.


In this work, we present a method to calculate the optical properties of III-V heterostructures based on a first principles approach. For this purpose, the band structure of the semiconductor heterostructure is obtained within $\mathbf{k\cdot p}$-theory~\cite{Kane1966} and the envelope function approach~\cite{Bastard1988} as described in \sref{sec:SingProp}. However, this approach heavily depends on material parameters such as effective masses which ultimately need to be extracted from experiments. To cope with this issue, we use density functional theory (DFT)~\cite{Kohn1965} to calculate the bulk band structures of each quantum well (QW) within the sample. To overcome the common shortcoming of DFT to underestimate the band gap of semiconductors~\cite{Perdew1983}, an advanced functional is used. A bulk $\mathbf{k\cdot p}$-band structure~\cite{Kane1966} is then fitted to its DFT counterpart. This approach results in a set of effective material parameters which are used to obtain the band structure of the QW sample within the envelope function approach.

The resulting energy bands and wave functions from the $\mathbf{k\cdot p}$-calculation serve as starting point for the calculation of the optical properties. We calculate the absorption using the semiconductor Bloch approach~\cite{Haug2004} as outlined in \sref{sec:OptProp}. The computation of other quantities such as the refractive index or photo luminescence is also possible within our microscopic approach. However, in this work we only aim at demonstrating the possibility of combining DFT and the semiconductor Bloch approach. Therefore, we chose a Ga(AsP) QW with (AlGa)As barriers, a well known III-V material system, for demonstration purposes.

It is worth noting that the $8\times8$ $\mathbf{k\cdot p}$-Hamiltonian applied in this work can be extended to include conduction band anti-crossing (CBAC) allowing the description of the band structure of dilute nitrides~\cite{Shan2001}. Furthermore, it can be extended to include valence band anti-crossing (VBAC) that allows to describe the band structure of alloys such as Ga(AsBi)~\cite{Alberi2007} or even both CBAC and VBAC for Ga(NAsBi) resulting in a $14\times 14$ Hamiltonian~\cite{Broderick2013} (10 valence and 4 conduction bands).

\section{Methods}

\subsection{DFT calculations}
All density functional theory calculations were performed with the Vienna \textit{ab initio} Simulation Package (VASP)~\cite{VASP1993, VASP1994, VASP1996a, VASP1996b}. The lattice constants of ternary alloys were calculated with Vegard's rule based on computed lattice constants of the binary constituents at the accuracy described below. These alloy lattice constants do not change during relaxation. Atomic positions were relaxed at the generalized gradient approximation level (GGA) with the PBE functional~\cite{PBE, PBE_Erratum}, including Grimme's D3(BJ) dispersion correction~\cite{GrimmeD3, GrimmeD3BJ}. The relaxation was performed with a SCF convergence of $10^{-7}\;\mathrm{eV}$, while forces were converged to $10^{-2}\;\mathrm{eV\cdot \mbox{\AA}^{-1}}$. Band structures were computed with the TB09 functional (also known as mBJLDA) which combines a meta-GGA exchange part with LDA correlation~\cite{Tran2009}. Here, eigenvalues were converged to $10^{-5}\;\mathrm{eV}$. A plane wave basis set with a kinetic energy cut-off of $450\;\mathrm{eV}$ was used in conjunction with the projector augmented wave (PAW) method~\cite{PAW_Blochl, PAW_Kresse}. Sampling of $k$-space was performed with a $\Gamma$-centered Monkhorst-Pack grid~\cite{monkhorstpack} using six points per direction for primitive zinc-blende type unit cells and two points per direction for 54 atom supercells. For band structure calculations spin-orbit coupling (SOC) was enabled and usage of all symmetries, including $k$-space, was turned off. 

Supercell (SC) band structures were projected onto the primitive cell (PC) Brillouin zone (``unfolded'') with BandUP, yielding an effective band structure (EBS)~\cite{Medeiros2014, Medeiros2015}. The EBS assigns a spectral weight $w(\mathbf{k}, \varepsilon)$ to PC-wavevectors and energies. This corresponds to the number of states at the given point and is a measure for the Bloch character. Since the translational symmetry of the primitive lattice is broken by substitutions and relaxations in the SC, the spectral weight can take non-integer values. The energy was resolved to $20\;\mathrm{meV}$ in the unfolding procedure.

As SCs we always used special quasirandom structures (SQS)~\cite{Zunger1990} with 54 atoms in this work. The SQS were computed with the \textit{Alloy-theoretic automated toolkit} (ATAT)~\cite{Walle2002,Walle2009,Walle2013} where we set the pair length to include third nearest neighbors, the triplet length to include second nearest neighbors, and the quadruplet length to include nearest neighbors.

\subsection{$\mathbf{k\cdot p}$ calculations\label{sec:SingProp}}
Starting point for our $\mathbf{k\cdot p}$ calculations is the fully coupled $8\times8$ bulk $\mathbf{k\cdot p}$-Hamiltonian~\cite{Hader1997}. 
Included are the twice spin-degenerate $s$-like conduction band, $p$-like heavy-hole, light-hole, and split-off bands as well as the influence of remote bands via \textit{L{\"o}wdin renormalization} of the Luttinger parameters $\gamma_1, \gamma_2, \gamma_3$, and $\gamma_c$~\cite{Lowdin1951, Cardona1988}. The set of renormalized Luttinger parameters and the Kane energy $E_\mathrm{P}$ for bulk materials are obtained from DFT as described in the next section. Since we are dealing with heterostructures, we employ the envelope function approach~\cite{Bastard1988, Winkler1993} to account for the quantum confinement of the charge carriers and include epitaxial strain using the \textit{Pikus-Bir} formalism~\cite{Bir1974,Chuang1991}. Diagonalizing the resulting Hamiltonian yields the energy dispersion and eigenvectors which are used to obtain the dipole and Coulomb matrix elements. All these are required as input for the calculation of the optical properties (see~\sref{sec:OptProp}).

\subsection{Extraction of $\mathbf{k\cdot p}$-parameters from DFT-band structures}
In contrast to the spectral weight of the EBS, diagonalizing the $\mathbf{k\cdot p}$-Hamiltonian will yield a set of discrete bands following a dispersion relation $\varepsilon(\mathbf{k})$. Thus, the EBS must first be reduced to a band structure of the same form. This is carried out in two steps. First, the EBS is averaged over in order to reduce the smearing of the effective bands. Averaging is performed in a small energy window. Since smearing tends to increase with larger distance to $\Gamma$ for many systems, the averaging can and sometimes needs to be performed over smaller windows close to $\Gamma$ and over larger windows further away. This leads to an averaged band structure as an intermediate which is less smeared out than the EBS, but still contains spectral weights with generally non-integer values. In the second step, the spectral weights of the averaged band structure are rounded to the next nearest integer and a band structure with an appropriate number of bands at a given wave-vector and energy is obtained. The relevant number of conduction and valence bands constitute the reduced band structure. This band structure has the desired form and can easily be compared to a $\mathbf{k\cdot p}$ calculation. A caveat should be mentioned: this procedure requires that the averaged spectral weight is sufficiently close to an integer value, meaning that the Bloch character is not reduced too strong. For example, in the treatment of dilute nitrides and bismides the described procedure needs to be modified because the band anti-crossing leads to sub-band states that can have a small spectral weight.

In order to compare the reduced band structure with its $\mathbf{k\cdot p}$ analog, a measure for the deviation between them must be established. This is achieved by summing over the squared difference of energies between both band structures, weighted by the distance to the $\Gamma$-point:
\begin{equation*}
\delta = \sum_{\mathbf{k}}^{\mathbf{k}_{\mathrm{cut}}} \sum_i^{N_{\mathrm{bands}}} (\varepsilon_{\mathrm{DFT}} - \varepsilon_{\mathbf{kp}})^2 \cdot \left(1 - \frac{\sqrt{\mathbf{k} \cdot \mathbf{k}}}{k_{\mathrm{cut}}} \right).
\end{equation*}
The weighting ensures that momenta close to the center of the Brillouin zone influence the outcome of the fit stronger resulting in a good agreement of the fit in this region. This is desirable since these $k$-points are more important for optical properties and are better described by $\mathbf{k\cdot p}$-theory than $k$-points further away from the center. Only $k$-points within a sphere with radius $k_{\mathrm{cut}}$ are taken into account for the fitting procedure. In this work, $k_{\mathrm{cut}} = 0.15\;\mathrm{\mbox{\AA}}^{-1}$.

The actual fit is performed by varying the starting parameters in a two-step procedure in order to reduce $\delta$. In the first step, the parameters are varied by a fixed relative amount (here: $10\;\%$ of the most recent value) until the difference of $\delta$ to the last step is smaller than $0.001\delta_0$ (with $\delta_0$ being $\delta$ for the initial parameters). A variation is accepted only if it leads to a smaller $\delta$. It turned out that this step is more stable if not all parameters are varied at the same time. Thus, at first only $\gamma_1^{\prime}$, $\gamma_2^{\prime}$, and $E_\mathrm{P}$ are varied, then $\gamma_3^{\prime}$, and finally $\gamma_\mathrm{c}^{\prime}$. Especially the last two parameters showed to be unstable if varied simultaneously.

In the second step, all parameters are varied by small random numbers. Again, new values are only accepted if they lead to a smaller $\delta$. This is performed for 2000 variations of each parameter which showed to lead to a well converged parameter set. Also, no instability of the solution was observed when all parameters were simultaneously changed during this second step.

\subsection{Calculation of optical properties\label{sec:OptProp}}

The absorption spectra are calculated using a microscopic theory which is based on the multi-band semiconductor Bloch equations~\cite{Lindberg1988,Girndt1997,Hader2003,Haug2004}. As input parameters single particle properties are obtained as described in \sref{sec:SingProp}. The equation of motion for the microscopic interband polarization $p_{\pmb{k}}^{\lambda\nu}$ between an electron in band $\lambda$ and a hole in band $\nu$ is 
\begin{eqnarray*}
 \frac{\textnormal{d}}{\textnormal{d}t}p^{\lambda\nu}_{\pmb{k}} &= \frac{1}{i\hbar}\Bigg[\sum_{\lambda',\nu',\pmb{k}}\big(\varepsilon^{\nu,\nu'}_{\pmb{k}} \delta_{\lambda,\lambda'}+\varepsilon^{\lambda,\lambda'}_{\pmb{k}}\delta_{\nu,\nu'}) p^{\lambda'\nu'}_{\pmb{k}}\nonumber\\*
 &+\left(1-f^{\lambda}_{\pmb{k}}-f^{\nu}_{\pmb{k}}\right)\Omega^{\lambda,\nu}_{\pmb{k}}\Bigg] + \frac{\textnormal{d}}{\textnormal{d}t}p^{\lambda\nu}_{\pmb{k}}\bigg|_{\textnormal{corr}}\, ,
\end{eqnarray*}
where $\varepsilon_{\pmb{k}}$ are the renormalized electron (superscript $\lambda,\lambda'$) and hole (superscript $\nu,\nu'$) energies, $f_{\pmb{k}}^{\lambda(\nu)}$ is the electron (hole) density, and $\Omega_{\pmb{k}}^{\lambda,\nu}$ is the renormalized field. The renormalization is due to the Coulomb interaction of charge carriers. Microscopic electron-electron and electron-phonon scattering that cause a dephasing of the polarization are included via $p^{\lambda\nu}_{\pmb{k}}\big|_{\textnormal{corr}}$. We treat these correlation terms in second order Born-Markov approximation~\cite{Haug2004}. The equation of motion for the polarization couples to the equations of motion for the electron and hole densities.
From the microscopic polarization the absorption and refractive index can be calculated.
For a more detailed explanation of the terms and methods, the reader is referred to~\cite{Girndt1997,Hader2003,Haug2004}. In this work, all absorption calculations are performed at $T = 300\,$K.

\section{Results and Discussion}

\subsection{Method validation}
The computational approach was tested on the well studied system GaAs, since it offers reliable reference values and can be computed at low computational cost. Furthermore, it is the host for most alloys of interest. The band structure was computed with the global hybrid functional PBE0~\cite{Perdew1996}, the range-separated hybrid functional HSE06~\cite{Krukau2006}, and the TB09 functional \cite{Tran2009}. The resulting data are listed in \tref{tab:kpgaas}. The band gap as an experimentally accessible parameter is well reproduced by TB09 with $1.44\;\mathrm{eV}$ compared to an experimental value of $1.52\;\mathrm{eV}$ at $0\,$K~\cite{Vurgaftman2001}.
In contrast, PBE0 yields an overestimated band gap of $\mathrm{1.71\;eV}$, while the range-separated HSE06 underestimates the gap stronger than TB09 ($\mathrm{1.11\;eV}$). A G$_0$W$_0$ calculation of the band gap yielded $\mathrm{1.41\;eV}$, showing that TB09 can reproduce features of much more demanding many-body methods, in accordance with other reported findings. 
Likewise, the $\Gamma$-point band gap of the minority component GaP could be reproduced with similar accuracy ($\mathrm{2.95\;eV}$ with TB09, $\mathrm{2.86\;eV}$ exp.~\cite{Vurgaftman2001}).

\begin{table*}
\caption{Band gap, spin-orbit splitting and effective $\mathbf{k\cdot p}$-parameters of GaAs determined with various functionals. The lattice constant of $5.68942\;\mathrm{\mbox{\AA}}$ was computed with PBE-D3.}\label{tab:kpgaas}
\begin{indented}
\lineup
\item[]\begin{tabular}{@{}llllllll@{}}
\br
 & $E_\mathrm{g}$ (eV) & $\Delta_{\mathrm{SO}}$ (eV) & $\gamma_1^{\prime}$  & $\gamma_2^{\prime}$ & $\gamma_3^{\prime}$ & $\gamma_\mathrm{c}^{\prime}$  & $E_\mathrm{P}$ (eV)\\ 
\mr
PBE0           & 1.71 & 0.38 & 4.238 & \-0.014 & 0.003  & \-0.301 & 59.546 \\
HSE06          & 1.11 & 0.37 & 4.456 & \-0.014 & 0.003  & \-0.293 & 50.541 \\
TB09           & 1.44 & 0.32 & 1.119 & \-0.657 & 0.017  & \-0.828 & 19.876 \\
TB09$^{a}$ & 1.52 & 0.32 & 1.130 & \-0.656 & 0.018  & \-0.813 & 20.661 \\
Ref.~\cite{Vurgaftman2001} & 1.52 & 0.34 & 1.13  & \-0.759 & 0.0405 & \-0.538 & 26.06  \\ 
\br
\end{tabular}
\item[] $^{a}$ After application of a scissor operator.
\end{indented}
\end{table*}

Applying the fitting routine to the computed band structure, the hybrid functionals deviate significantly -- and consistently with each other -- from the reference values for the $\mathbf{k\cdot p}$-parameters, indicating that neither reproduces the band curvature sufficiently well. TB09 performs better overall, showing its most significant deviations for the $\gamma_3^{\prime}$ and $\gamma_\mathrm{c}^{\prime}$ parameters. The latter directly influences the curvature of the conduction bands which are notoriously difficult to describe with DFT, while the former shows up in matrix elements coupling valence states with each other. The curvature of all bands is influenced rather strongly by the Kane energy $E_\mathrm{P}$, which may compensate the deviation of other parameters. 

Shifting the conduction band to the experimental band gap by applying a scissor operator prior to fitting only has a small influence on the $\mathbf{k\cdot p}$-parameters. Overall, the agreement to the reference values is improved slightly. Applying a scissor operator at this point may be desirable, since the band gap directly controls optical properties like the onset of absorption and the photo luminescence wavelength. The agreement between ``scissored'' and ``unscissored'' band parameters shows that this step can be applied without changing the result of the fit.

\subsection{Determination of $\mathbf{k\cdot p}$-parameters for alloys}\label{sec:kp-paras}
\begin{figure}
\centering
\includegraphics[width=15.7cm]{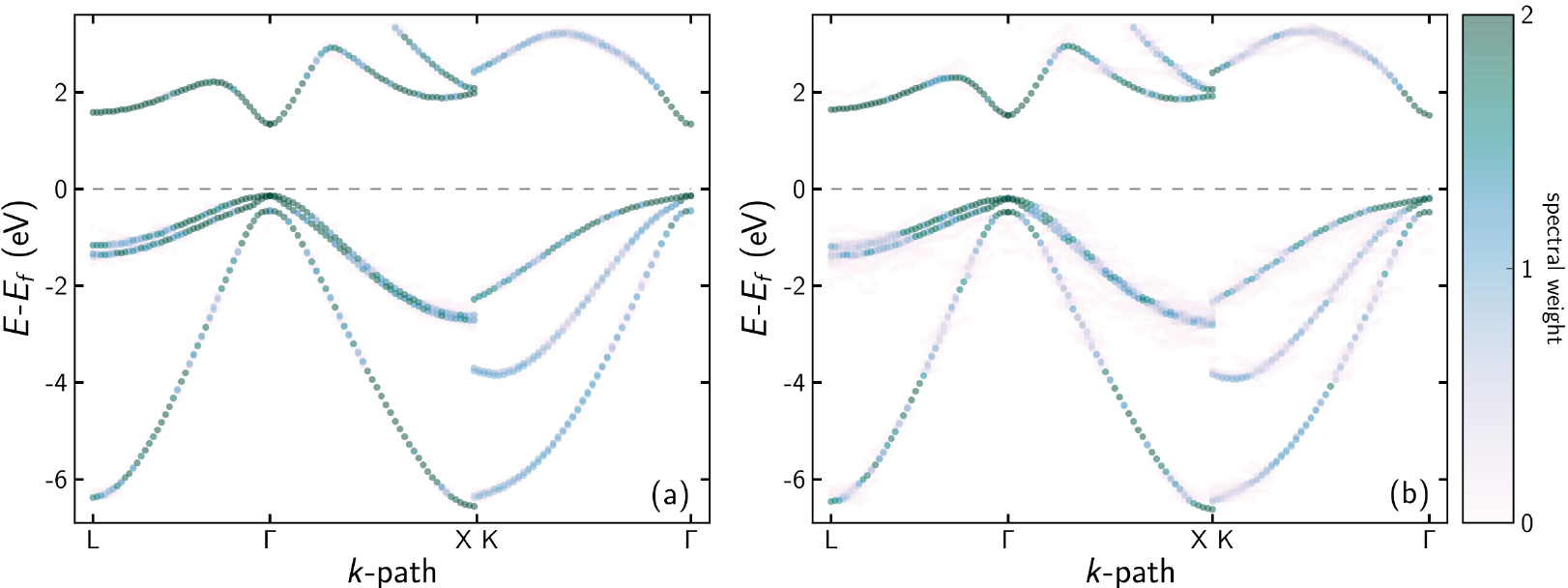}
\caption{Effective band structure (EBS) relative to the Fermi level $E_f$ of (a) $\mathrm{Ga_{27}(As_{26}P_1)}$ ($x_\mathrm{P} = 3.7\;\%$) and (b) $\mathrm{Ga_{27}(As_{21}P_{6})}$ ($x_\mathrm{P} = 22.2\;\%$). The color gradient reflects the spectral weight $0 \leq w(\mathbf{k}, \varepsilon) \leq 2$. For both band structures, states with a spectral weight $< 0.01$ were omitted for better visibility.}\label{fig:ebs-gapas}
\end{figure}

For all DFT band structure calculations of ternary compounds we used SQS. The EBS of Ga$_{27}$(As$_{26}$P$_1$) ($x_\mathrm{P} = 3.7\;\%$) and of Ga$_{27}$(As$_{21}$P$_6$) ($x_\mathrm{P} = 22.2\;\%$), which are the extremes of the concentration range studied here, are shown in \fref{fig:ebs-gapas}. In both cases, a strong Bloch character is retained. Some breaking of the translational symmetry is reflected by the ``smearing'' of the bands. This increases with the P concentration due to enhanced symmetry breaking. Especially in the vicinity of the $\Gamma$-point, which is most important for the optical properties, the smearing is small and a high Bloch character prevails. Visible smearing occurs at the Brillouin zone edge and especially at the folding lines, i.~e. at $1/3$ and $2/3$ of the path to the edge.

\begin{figure}
\centering
\includegraphics[width=7.7cm]{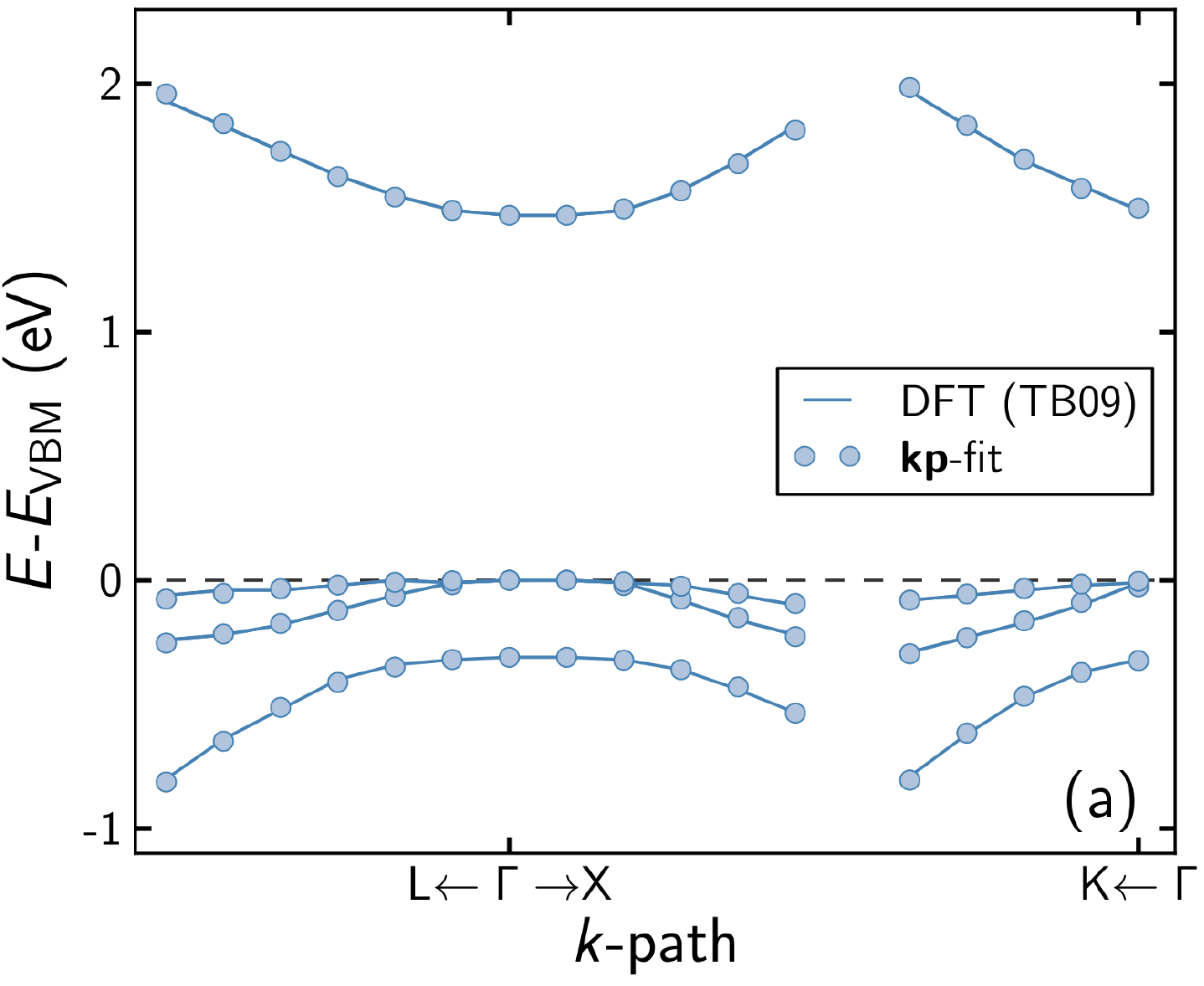}\hspace{8pt}\includegraphics[width=7.7cm]{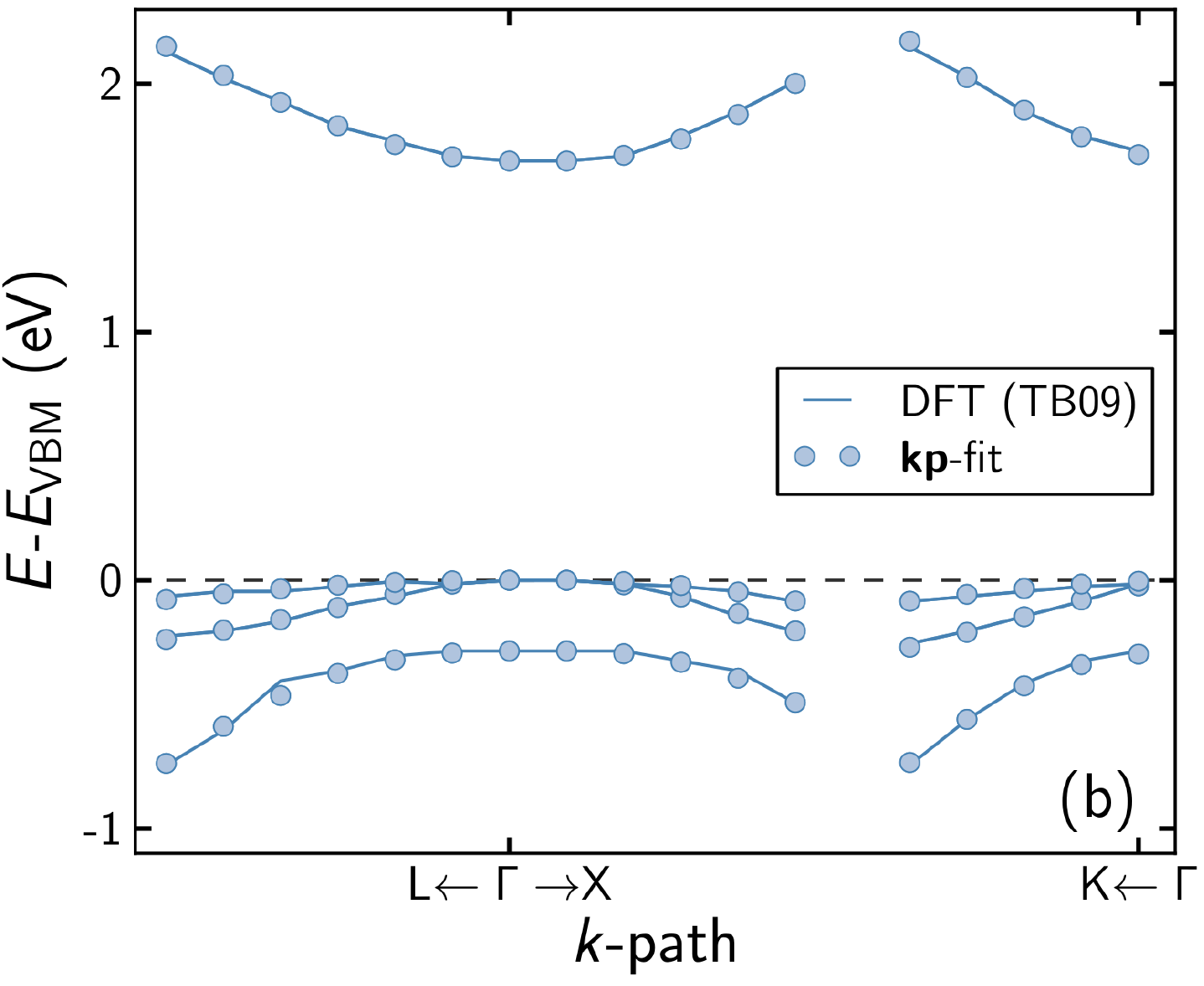}
\caption{Reduced DFT-band structure and result of $\mathbf{k\cdot p}$-fit relative to the valence band maximum ($E_\mathrm{VBM}$) of (a) Ga$_{27}$(As$_{26}$P$_1$) and (b) Ga$_{27}$(As$_{21}$P$_6$) within the region of $k$-space used for fitting $|k| \leq 0.15\,$\AA$^{-1}$.}\label{fig:fit-gapas}
\end{figure}

Averaging over these EBS and reducing them to eight effective bands is straightforward.
The reduced band structures are then used for fitting to extract the $\mathbf{k\cdot p}$-parameters. The resulting parameters reproduce the DFT-band structure within the fitting range well (see~\fref{fig:fit-gapas}) and are listed in \tref{tab:kpgapas}. While the band gap increases consistently with the P concentration (and the spin-orbit splitting decreases), no clear trend can be assigned to any of the other parameters. 

The DFT calculations are at $T = 0\,$K whereas we want to calculate the optical properties at $300\,$K. In order to obtain band structures at $T = 300\,$K
we apply a scissor operator to shift the band gaps to values obtained from a temperature dependent virtual crystal model~\cite{Vurgaftman2001} before fitting. 
 This only weakly affects the band parameters obtained by fitting, keeping the lack of a clear trend and the strong scattering of values (see~\tref{tab:kpgapas-scissored}). Again, the scissor operator can be safely applied without changing the characteristics of the $\mathbf{k\cdot p}$-bands. 

An alternative approach to the virtual crystal model is to calculate the band gaps of the ternary alloys using DFT. The only information we require is the experimental band gap of GaAs at the respective temperature. By variation of the parameter $c$ in the TB09 functional~\cite{Tran2009} as done by Kim~\etal~\cite{Kim2010} it is possible to accurately fit the GaAs band gap. When using the original TB09 implementation, $c$ is calculated from the electron density and its gradient~\cite{Tran2009}. Here, we calculate the band gap of GaAs with the original $c$ and then find a $c'$ that reproduces the experimental GaAs band gap. We then use the difference $\Delta c = c'-c$ as a correction in the calculation of the band gap of the ternary compound Ga(AsP): First, we calculate the band gap with the original TB09 implementation and then we correct the respective value of $c$ by $\Delta c$. The band gaps calculated with the corrected $c'$ reproduce the band gaps from the virtual crystal model within a range of $\pm\,10\,$meV for all concentrations considered here.

\begin{table*}
\caption{Band gap, spin-orbit splitting and $\mathbf{k\cdot p}$-parameters for the binary compounds GaAs and GaP as well as Ga$_{27}$(As$_{27-x}$P$_{x}$) with $x = 1-6$ and (Al$_{9}$Ga$_{18}$)As$_{27}$ SQS-supercells derived from effective band structures computed with the TB09 functional. The lattice constants $a_0$ of the binary compounds were determined with PBE-D3 and of the ternary compounds from the binary lattice constants using Vegard's rule.}
\label{tab:kpgapas}
\begin{indented}
\lineup
\item[]\begin{tabular}{@{}lllllllll}
\br
 & $E_\mathrm{g}^{(a)}$ & $\Delta_{\mathrm{SO}}^{(a)}$ & $\gamma_1^{\prime}$  & $\gamma_2^{\prime}$ & $\gamma_3^{\prime}$ & $\gamma_\mathrm{c}^{\prime}$  & $E_\mathrm{P}^{(a)}$ & $a_0\,$(\AA)\\
\mr
GaAs                        & 1.44 & 0.32 & 1.087 & \-0.673 & 0.002  & \-0.876 & 20.022 & 5.68942  \\
GaP                         & 2.95 & 0.08 & 1.150 & \-0.479 & 0.109  & \-0.628 & 19.875 & 5.47747  \\[.4ex]
Ga(As$_{26}$P$_1$)          & 1.48 & 0.31 & 1.167 & \-0.633 & 0.001 & \-0.877 & 19.884 & 5.68157 \\
Ga(As$_{25}$P$_2$)          & 1.52 & 0.30 & 1.021 & \-0.649 & 0.001 & \-0.994 & 19.876 & 5.67371 \\
Ga(As$_{24}$P$_3$)          & 1.59 & 0.29 & 0.919 & \-0.755 & 0.016 & \-1.055 & 20.025 & 5.66586 \\
Ga(As$_{23}$P$_4$)          & 1.62 & 0.28 & 1.124 & \-0.639 & 0.055 & \-0.750 & 19.817 & 5.65810 \\
Ga(As$_{22}$P$_5$)          & 1.68 & 0.28 & 0.972 & \-0.690 & 0.007 & \-1.020 & 20.229 & 5.65016 \\
Ga(As$_{21}$P$_6$)          & 1.72 & 0.28 & 1.186 & \-0.478 & 0.000 & \-0.552 & 19.026 & 5.64230 \\[.4ex]
(Al$_{9}$Ga$_{18}$)As       & 1.86 & 0.30 & 0.792 & \-0.640 & 0.001 & \-1.046 & 19.492 & 5.68219  \\
\br 
\end{tabular}
\item[] $^{(a)}$ Units in eV.
\end{indented}
\end{table*}

\begin{table*}
\caption{Band gap, spin-orbit splitting and $\mathbf{k\cdot p}$-parameters for the binary compounds GaAs and GaP as well as Ga$_{27}$(As$_{27-x}$P$_{x}$) with $x = 1-6$ and (Al$_{9}$Ga$_{18}$)As$_{27}$ SQS-supercells derived from effective band structures computed with the TB09 functional. A scissor operator was applied prior to fitting to shift the band gap to the reference value at $T=300\,$K~\cite{Vurgaftman2001}.}
\label{tab:kpgapas-scissored}
\begin{indented}
\lineup
\item[]\begin{tabular}{@{}lllllllll}
\br
 & $E_\mathrm{g}^{(a)}$ & Scissor$^{(b)}$ & $\Delta_{\mathrm{SO}}^{(a)}$ & $\gamma_1^{\prime}$  & $\gamma_2^{\prime}$ & $\gamma_3^{\prime}$ & $\gamma_\mathrm{c}^{\prime}$  & $E_\mathrm{P}^{(a)}$\\
\mr
GaAs                               & 1.42 & \-0.02 & 0.32 & 1.087 & \-0.672 & 0.003 & \-0.877 & 19.805  \\
GaP                                & 2.78 & \-0.17 & 0.08 & 1.048 & \-0.530 & 0.059 & \-0.781 & 19.657  \\[.4ex]
Ga$_{27}$(As$_{26}$P$_1$)          & 1.47 & \-0.01 & 0.31 & 1.168 & \-0.632 & 0.001 & \-0.877 & 19.780  \\
Ga$_{27}$(As$_{25}$P$_2$)          & 1.51 & \-0.01 & 0.30 & 1.018 & \-0.649 & 0.001 & \-0.998 & 19.604  \\
Ga$_{27}$(As$_{24}$P$_3$)          & 1.55 & \-0.04 & 0.29 & 0.929 & \-0.750 & 0.023 & \-1.040 & 19.594  \\
Ga$_{27}$(As$_{23}$P$_4$)          & 1.60 & \-0.02 & 0.28 & 1.127 & \-0.637 & 0.058 & \-0.745 & 19.579  \\
Ga$_{27}$(As$_{22}$P$_5$)          & 1.64 & \-0.04 & 0.28 & 0.965 & \-0.692 & 0.004 & \-1.030 & 19.899  \\
Ga$_{27}$(As$_{21}$P$_6$)          & 1.69 & \-0.03 & 0.28 & 1.184 & \-0.478 & 0.000 & \-0.554 & 18.800  \\[.4ex]
(Al$_{9}$Ga$_{18}$)As$_{27}$       & 1.88 & \phantom{\-}0.02 & 0.30 & 0.792 & \-0.641 & 0.001 & \-1.046 & 19.674  \\
\br 
\end{tabular}
\item[] $^{(a)}$ Units in eV.
\item[] $^{(b)}$ Scissor shift applied on $E_g$ from \tref{tab:kpgapas}. Units in eV.
\end{indented}
\end{table*}

\begin{figure}
\centering
\includegraphics[width=6.8cm]{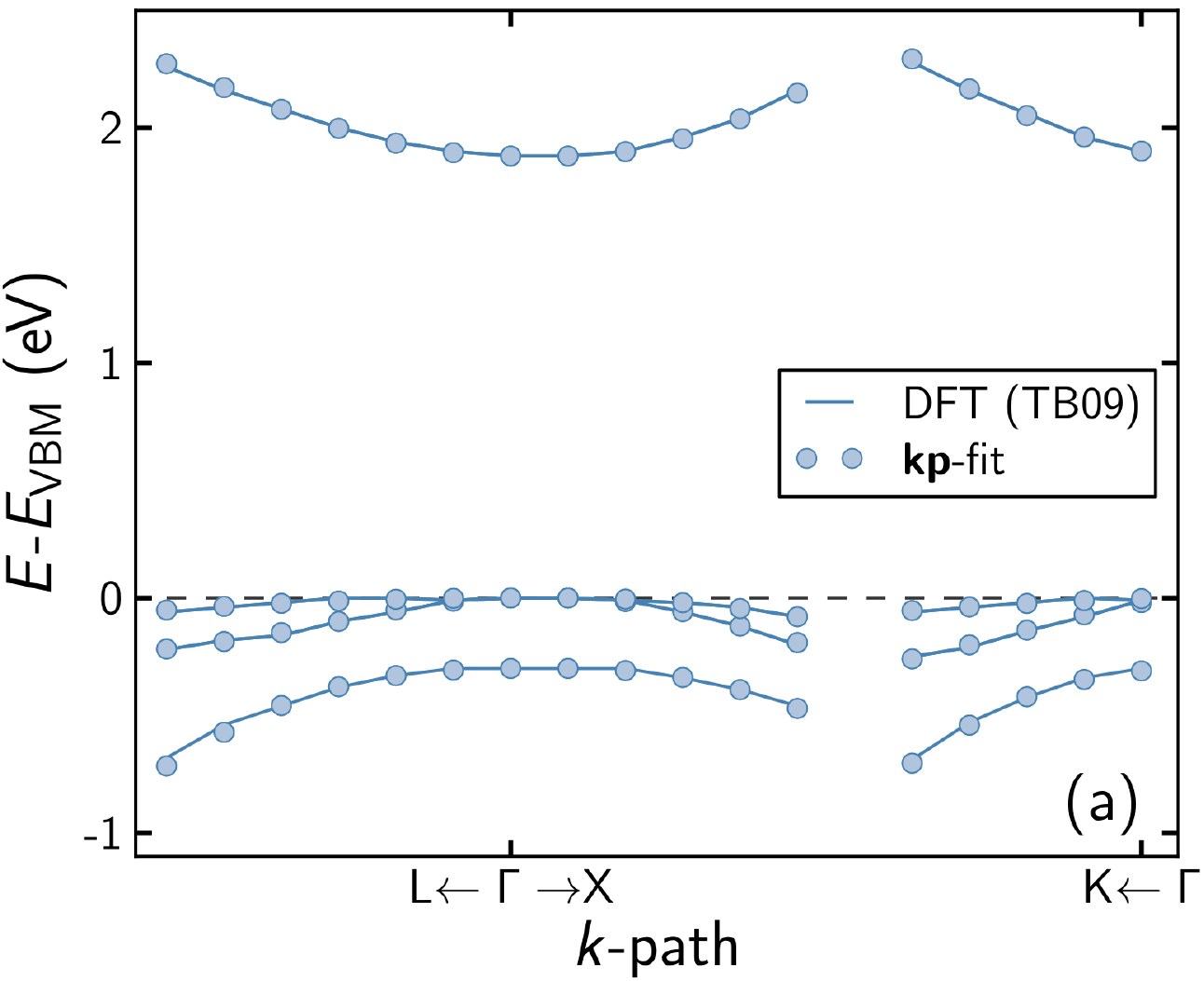}\hspace{5.0pt}\includegraphics[width=8.3cm]{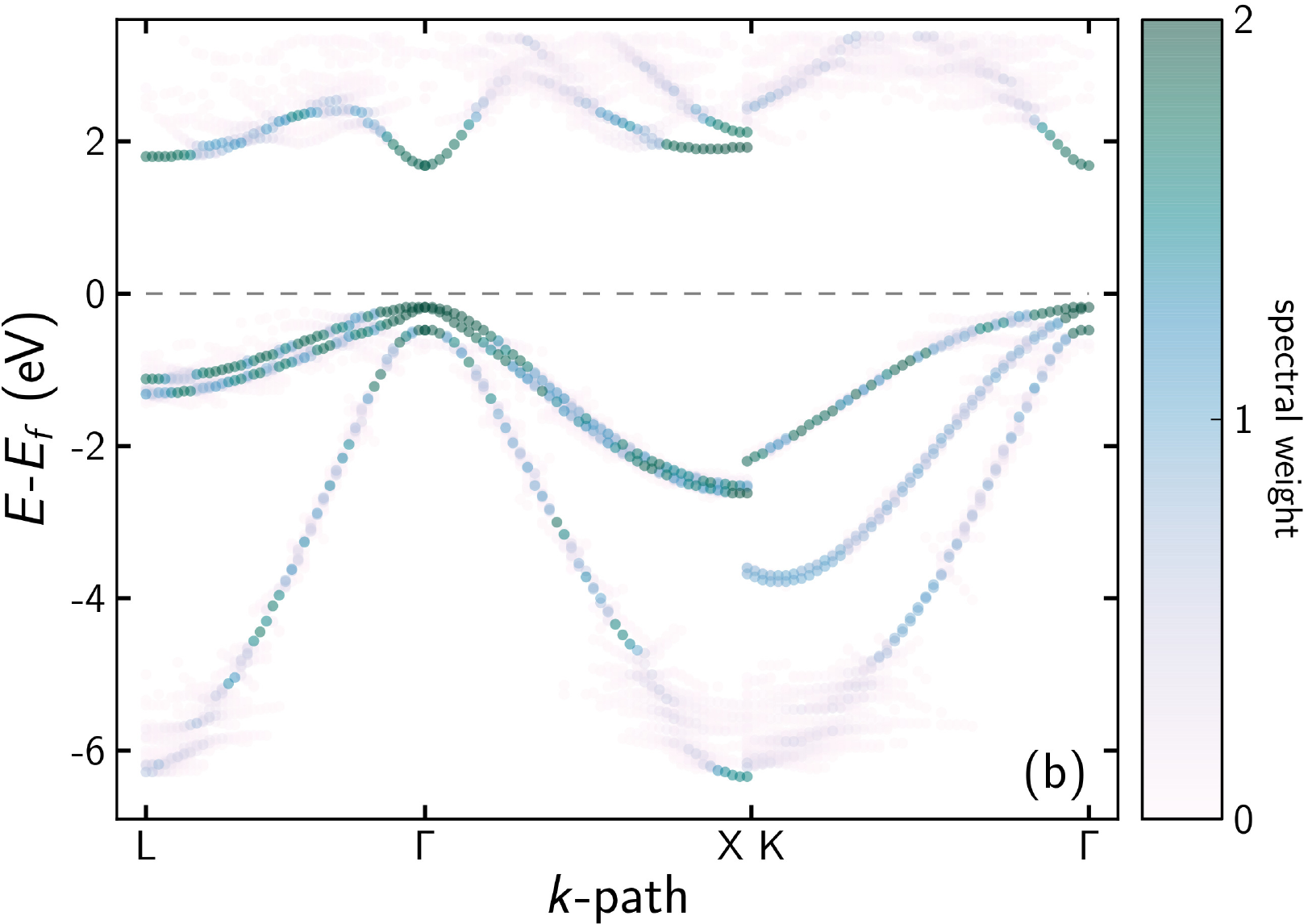}
\caption{(a) Reduced DFT-band structure and $\mathbf{k\cdot p}$-fit of (Al$_{18}$Ga$_9$)As$_{27}$ relative to the valence band maximum ($E_\mathrm{VBM}$) and (b) EBS of (Al$_{18}$Ga$_9$)As$_{27}$ relative to the Fermi level $E_f$. In (a), the DFT and $\mathbf{k\cdot p}$-band structure are shown for the region of $k$-space used for fitting $|k| \leq 0.15\,$\AA$^{-1}$. In (b) the color gradient reflects the spectral weight $0 \leq w(\mathbf{k}, \varepsilon) \leq 2$. States with a spectral weight $w(\mathbf{k}, \varepsilon) < 0.01$ were omitted for better visibility.}
\label{fig:ebs-algaas}
\end{figure}

The EBS and fitting results of the barrier material (Al$_{9}$Ga$_{18}$)As$_{27}$ are shown in \fref{fig:ebs-algaas}. While the valence band states in \fref{fig:ebs-algaas}(b) mostly retain a strong Bloch character similar to the dilute phosphides, the conduction band states are strongly smeared, an exception being the lowest ones close to the $\Gamma$-point. This is exactly the region required for fitting, so averaging the EBS and reducing it to eight bands was again straightforward.
The resulting parameters are shown in \tref{tab:kpgapas} and \tref{tab:kpgapas-scissored} without and with scissor shift, respectively. As for the other cases described in this work, the $\mathbf{k\cdot p}$-parameters are hardly affected by the scissor shift.

\begin{figure}
\centering
\includegraphics[width=8.32cm]{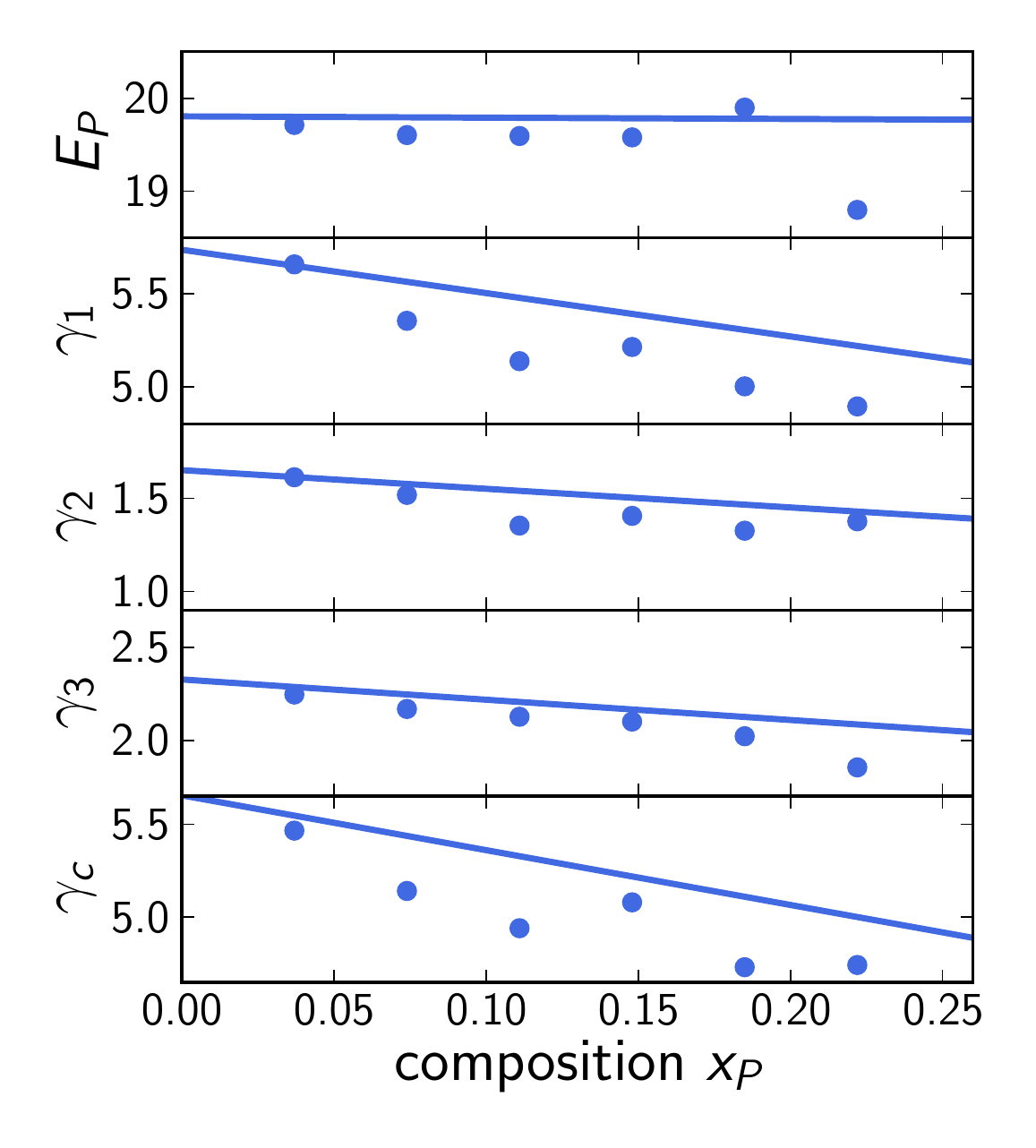}
\caption{Kane energy $E_P$ and Luttinger parameters $\gamma$ of the binary compounds GaAs and GaP were interpolated (solid blue lines) and are compared to the $\mathbf{k\cdot p}$-parameters obtained from ternary Ga(AsP)-SQS calculations (blue dots).}\label{fig:luttpara}
\end{figure}

In addition to the $\mathbf{k\cdot p}$-parameters obtained from fitting the ternary Ga(AsP)-SQS band structures, we calculated $\mathbf{k\cdot p}$-parameters for Ga(AsP) by linearly interpolating those of the binary compounds GaAs and GaP from \tref{tab:kpgapas-scissored}. Interpolation was performed for the Kane energies $E_P$ and un-renormalized Luttinger parameters $\gamma$. The latter are connected to the renormalized ones $\gamma'$ listed in \tref{tab:kpgapas-scissored} via (cf.~\cite{Lowdin1951,Cardona1988})
\begin{eqnarray*}
 \gamma_1 = \gamma_1' +\frac{1}{3}\frac{E_P}{E_g}\,,\\
 \gamma_2 = \gamma_2' +\frac{1}{6}\frac{E_P}{E_g}\,,\\
 \gamma_3 = \gamma_3' +\frac{1}{6}\frac{E_P}{E_g}\,,\\
 \textnormal{and}\\
 \gamma_c = \gamma_c' +\frac{1}{3}\left(\frac{E_P}{E_g}+\frac{1}{2}\frac{E_P}{E_g+\Delta_{SO}}\right)\,.
\end{eqnarray*}
The comparison of the un-renormalized Luttinger parameters and the Kane energies $E_P$ is shown in \fref{fig:luttpara}. It can be seen that the linearly interpolated $\mathbf{k\cdot p}$-parameters generally overestimate the SQS $\mathbf{k\cdot p}$-parameters. However, a very good agreement is obtained for $x_\mathrm{P} = 3.7\,$\%. 

\subsection{Quantum well optical properties}
To demonstrate our approach, we calculate the absorption spectra of a (AlGa)As/Ga(AsP)/(AlGa)As QW structure. The barrier material is chosen to be Al$_{0.33}$Ga$_{0.67}$As while for the $4\,$nm QW material we consider six different Ga(AsP) compositions by varying the P content $x_\mathrm{P}$ from  $3.7\,$\% to $22.2\,$\%.
From the parameter sets obtained by fitting the bulk DFT-band structures (see~\tref{tab:kpgapas-scissored}) the energy dispersion and wave functions for the eight spin degenerate bands of the QW are calculated (see \sref{sec:SingProp}). These serve as input for the computation of the absorption spectrum (see~\sref{sec:OptProp}) at $T=300\,$K.

For each phosphide composition two sets of absorption calculations were performed. We used the $\mathbf{k\cdot p}$-parameters of the Ga(AsP) SQS from \tref{tab:kpgapas-scissored} and we used interpolated $\mathbf{k\cdot p}$-parameters that were obtained as described in the last paragraph of \sref{sec:kp-paras} to set up the $\mathbf{k\cdot p}$-Hamiltonian. 

\begin{figure}
\centering
\includegraphics[width=7.7cm]{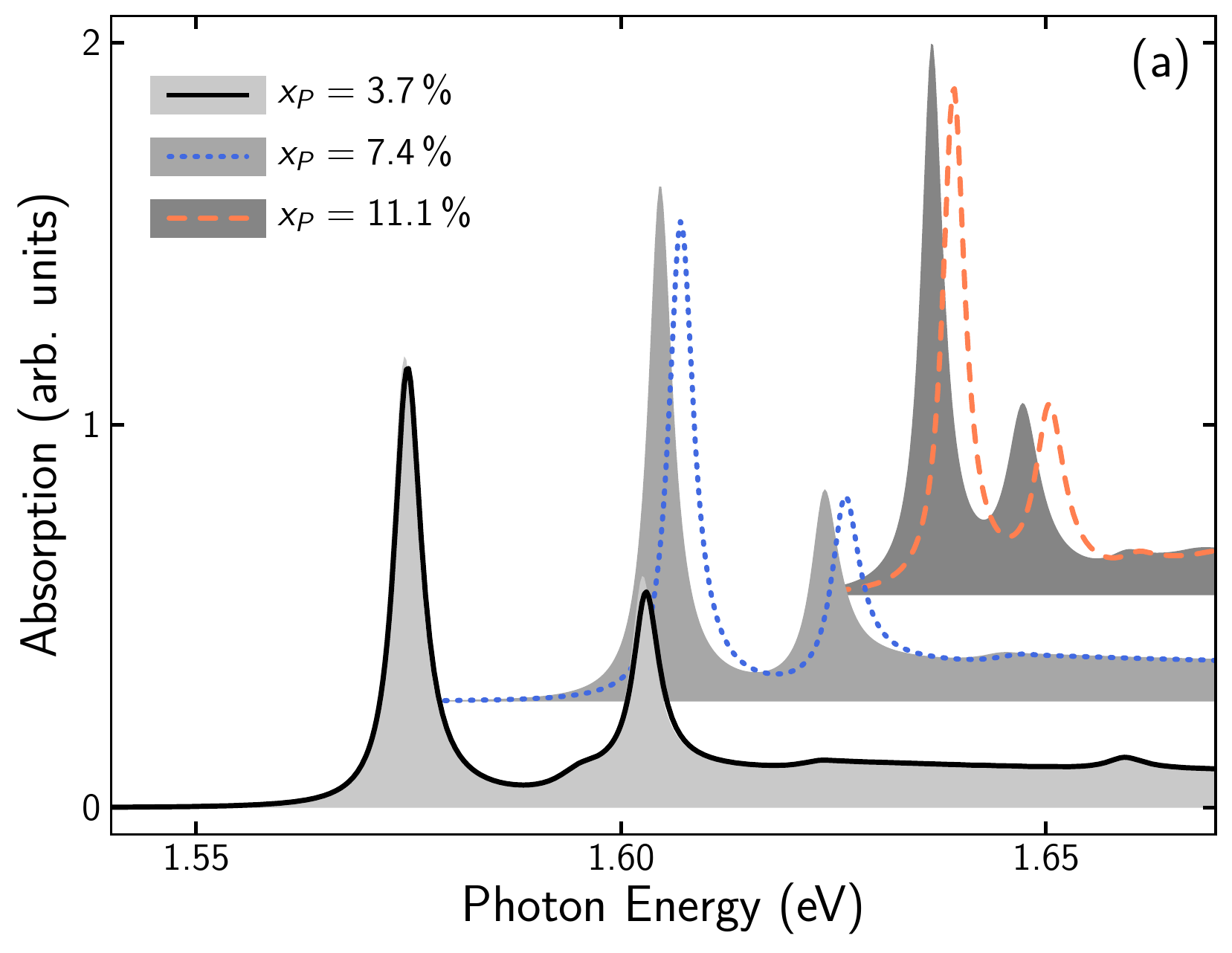}\hspace{8pt}\includegraphics[width=7.7cm]{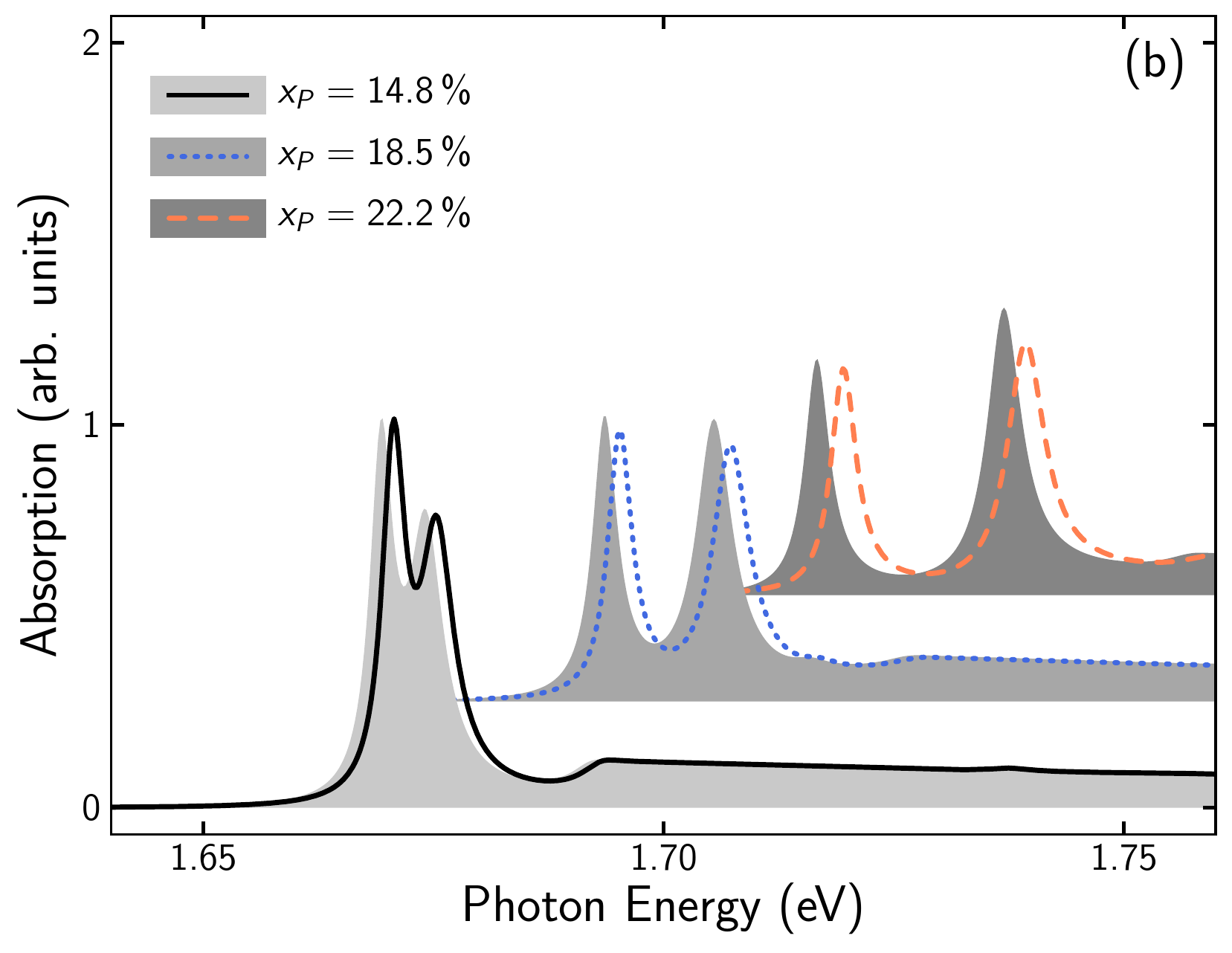}
\caption{Calculated absorption spectra as obtained from ternary Ga(AsP)-SQS $\mathbf{k\cdot p}$-parameters (grey areas) are compared to absorption spectra obtained from interpolating the $\mathbf{k\cdot p}$-parameters of the binary alloys GaAs and GaP (colored lines). In (a) the spectra for the phosphor compositions $x_\mathrm{P}$ = $3.7\,$\%$ - 11.1\,$\% are shown, in (b) the spectra for the phosphor compositions $x_\mathrm{P}$ = $14.8\,$\%$ - 22.2\,$\%.}\label{fig:absorption}
\end{figure}

In \fref{fig:absorption} we present the comparison between the absorption spectra obtained from calculations with linearly interpolated $\mathbf{k\cdot p}$-parameters (lines) and calculations using SQS $\mathbf{k\cdot p}$-parameters (grey areas). Overall, the two sets of calculations deliver comparable results and the shape of the absorption spectra of both sets of calculations coincides. However, in case of $x_\mathrm{P} = 7.4\,$\%, $x_\mathrm{P} = 11.1\,$\%, $x_\mathrm{P} = 18.5\,$\%, and $x_\mathrm{P} = 22.2\,$\% the spectra from the interpolated calculations are shifted slightly to higher energies by about $3\,$meV and the peak absorption in the spectra from the SQS calculations is larger. The good agreement of the absorption spectra between both sets of calculations at $x_\mathrm{P} = 3.7\,$\% and $x_\mathrm{P} = 14.8\,$\% is consistent with a good agreement of the input Luttinger parameters (see~\fref{fig:luttpara}).

In general, good agreements are expected for very dilute concentrations. Higher concentrations increase the disorder which results in a smearing of states in the EBS. This in turn leads to a higher inaccuracy of the fitted $\mathbf{k\cdot p}$-parameters. 

\section{Conclusion}
We have presented a new method to calculate optical properties of semiconductor heterostructures from first principles using a combination of $\mathbf{k\cdot p}$-theory, density functional theory, and the semiconductor Bloch approach. We demonstrate the applicability of the method by calculating the absorption spectra of (AlGa)As/Ga(AsP)(AlGa)As QWs where the phosphide concentration is varied between $3.7\,\% - 22.2\,\%$. 

The only external data we use is the temperature dependent band gap of the alloys. However, we also show that the band gap of Ga(AsP) at the concentrations considered here can be calculated with DFT very accurately using the TB09 functional. 

We also show that by interpolation of the Luttinger parameters from the binary compounds GaAs and GaP the absorption spectra obtained are comparable to those calculated for the truly ternary GaAs$_{1-x}$P$_x$ compounds. Both spectra match very well for small concentrations of phosphide. Adding more phosphide increases disorder effects and thus the smearing of the effective band structure.

Our method can, in principle, also be used to describe more complex compounds, e.g. quaternaries, dilute nitrides, and dilute bismides. The difficulty lies in fitting a disordered DFT band structure. For example, the incorporation of nitride strongly perturbs the conduction band while the incorporation of bismide into a host material mainly influences the valence bands. However, at dilute concentrations the distortion is small so a treatment within the here proposed method seems possible.

\ack
This work was funded by the DFG via the GRK 1782 ``Functionalization of Semiconductors''. LB, PR, EF, RT, and SWK thank the HRZ Marburg, CSC Frankfurt, and HLRS Stuttgart for providing computing time. The work at NLCSTR was supported by the Air Force Office of Scientific Research under the STR Phase II grant FA9550-16-C-0021. 

LCB and PR contributed equally to this work.\\

\bibliography{literatur}
\bibliographystyle{iopart-num}

\end{document}